\documentstyle[psfig]{mn}

\def\farcs{\hbox{$.\!\!^{\prime\prime}$}}

\title[Three new radio-loud quasars at z$>$4]{A search for distant 
radio-loud quasars in the CLASS survey: three new radio-selected quasars
 at z$>$4}

\author[I. Snellen et al.]
{\parbox[]{6.in}{I.A.G. Snellen$^{1,2}$, R.G. McMahon$^{1}$, 
 J. Dennett-Thorpe$^{3}$, N. Jackson$^4$, \\
K.-H. Mack$^{5,6,7}$, E. Xanthopoulos$^4$\\ 
\footnotesize
$^1$Institute of Astronomy, Madingley Road, Cambridge CB3 0HA, 
     United Kingdom\\
$^2$Institute for Astronomy, Royal Observatory, Blackford Hill,
Edinburgh EH9 3HJ, United Kingdom\\ 
$^3$Kapteyn Institute, Rijksuniversiteit Groningen, 9700 AV Groningen,      The Netherlands\\
$^4$Jodrell Bank Observatory, University of Manchester, Macclesfield, Cheshire 
     SK11 9DL, United Kingdom\\
$^5$Istituto di Radioastronomia de CNR, Via Gobetti 101, I-40129 Bologna, 
     Italy\\
$^6$Radioastronomisches Institut der Universit\"at Bonn, Auf dem H\"ugel 71,
D-53121 Bonn, Germany\\
$^7$ASTRON, P.O. Box 2, 7990 AA Dwingeloo, The Netherlands\\
}}

\date{}

\begin{document}
\maketitle

\begin{abstract}
We report on the search for distant radio-loud quasars in the
Cosmic Lens All Sky Survey (CLASS) of flat spectrum 
radio sources with $S_{5GHz}>30$ mJy. 
Unresolved optical counterparts were selected from APM scans of 
POSS-I plates, with $e<19.0$ and red
$o-e>2.0$ colours, in an effective area of $\sim6400$ deg$^{2}$. 
Four sources were found to be  quasars with $z>4$, 
of which one was previously known.
This sample bridges the gap between the strong radio surveys with
$S_{5GHz}>200$ mJy and the samples of radio-weak quasars that can be 
generated via radio observations of optically selected quasars.
 In addition, 4 new quasars at $z>3$ have been 
found. The selection criteria result
in a success-rate of $\sim$1:7 for radio-loud quasars at $z>4$, which is 
a significant improvement over previous studies.
This search yields a surface density of 1 per 1600 deg$^{2}$, which 
is about a factor of $\sim$15 lower than that found in a similar search
for radio-quiet quasars at $z>4$. The study presented here is 
strongly biased against quasars beyond $z>4.5$, since the $e$-passband of the 
POSS-I only samples the spectra shortward of 1200$\AA$ at these 
redshifts. 
\end{abstract}

\section{Introduction}

The abundance of massive compact objects at very early cosmological 
epochs firmly constrains cosmological structure formation scenarios 
(Kauffmann \& Haehnelt 2000, Efstathiou \& Rees 1988).
This stimulates the search for quasars at the highest possible redshifts.
Radio and optical surveys have shown a pronounced increase in the co-moving 
number density of quasars out to a redshift of 1 -- 2, by about two
orders of magnitude. There is now  evidence that at $z>2-3$
the abundance of quasars  declines
(Warren et al. 1994; Schmidt et al. 1995; Shaver et al. 1999), 
which may indicate that, at these 
redshifts, we witness the onset of quasar activity in the universe.
However, recently, the evidence for this has been debated by
Jarvis \& Rawlings (2000).
 
Whilst radio-loud quasars form only a small subset of the quasar population,
the construction of radio-loud quasar samples is less prone to selection 
effects than are optical samples, since radio emission is unaffected 
by either intrinsic or extrinsic absorption due to dust.
The most comprehensive and complete search for high redshift 
radio-loud quasars has been performed by Hook et al. (1995, 1996), who
observed  a complete sample of 161 $S_{5GHz}>200$ mJy flat spectrum 
radio sources in an area of 7300 deg$^{2}$, which were selected on their red 
optical colours and unresolved 
optical counterparts. Of these objects, 13 were found to be radio-loud quasars 
at $z>3$, from which one was located at $z=4.3$.
In a similar fashion, Hook et al. (1998) searched for high redshift radio-loud 
quasars among flat spectrum radio sources, about an order of magnitude
fainter ($S_{5GHz}>25$ mJy) than those  studied previously.
Seventy-three objects were selected from a complete sample of 2902 
radio sources, drawn from the Greenbank 5 GHz survey and the FIRST
1.4 GHz (Gregory \& Condon 1991; White et al. 1997) in an area of 
1600 deg$^{2}$ with $m_e<19.5$, of which 70\% 
were spectroscopically observed. Six of those were found to be 
radio loud quasars at $z>3$, resulting in a surface density of one
per 190 square degrees.

In this paper we report on a search for distant quasars in the
Cosmic Lens All Sky Survey (CLASS, Myers et al. 2001, in prep.), using
an optical colour selection technique similar, but more stringent, to that used
by Hook et al. (1995, 1996, 1998). This project was aimed to find 
radio-loud quasars at $z>4$, using candidate objects
at similar flux densities as those in the GB/FIRST sample, but over a larger
area.
At the time of writing, only 5 radio-selected quasars at $z>4$ are available 
in the literature. These are given in table \ref{litgt4}, including the 
newly discovered objects described in this paper.
In addition to these radio-loud quasars, radio galaxies have also 
been discovered at high redshift. These are
 selected on their ultra-steep radio spectrum.
At this moment, five radio galaxies are known at $z>4$ (e.g. van Breugel 2000).

\section{The Cosmic Lens All Sky Survey (CLASS)}

The CLASS survey is a large-scale survey primarily to search 
for gravitational lenses among flat-spectrum radio sources (Myers et al.,
2001, in preparation).
It consists of about $\sim 15,000$ flat spectrum sources, selected primarily 
from the  87GB 5 GHz, WENSS 325 MHz, and Greenbank 1400 MHz radio surveys
(Gregory \& Condon 1991; Rengelink et al. 1997; Condon \& Broderick 1986).
Recently a complete subsample of  11,685 objects was defined on the more 
accurate GB6 5 GHz (Gregory et al. 1996) and NVSS 1.4 GHz (Condon et al. 1998)
surveys. In the selection procedure, the 1.4 GHz flux density of a GB6 radio source is defined as all the NVSS flux within 70$''$ of the GB6 position.
 The complete sample has the following selection criteria:

\begin{itemize}
\item[1] Declination, $0^\circ<\delta<75^\circ$, and \\ 
  galactic latitude,  $|b|>10^\circ$.
\item[2] $\alpha_{GB6-NVSS}>-0.5$, with $S(\nu) \propto \nu^\alpha$.
\item[3] $S_{GB6} \ge 30$ mJy
\end{itemize}

At the time of the selection of the targets for our spectroscopy observations,
10170 objects ($\sim 87\%$) of this complete sample had been observed with 
the VLA in A configuration at 8.4 GHz, yielding a resolution of 0\farcs25
(fwhm) and
a position accuracy of $<0\farcs1$. The remaining $\sim1500$ objects, were
not yet observed due to gaps in the sky-coverage of the NVSS at that time. 
These may therefore be considered to be sources with random radio properties.

About 10\% of the observed targets were not detected at a $S_{8.4GHz}>5$ mJy 
level. About 30\% of those are certainly missed because they are 
 lobes of larger sources, 
which are  resolved out due to the lack of short baselines of the VLA
in A configuration. Another 10\% have been missed due to instrumental 
problems. Why the remaining sources have not been detected is not yet clear,
but extended or complicated structure may play a role. 
Hence, we believe that most of these objects have large angular sizes and 
are  unlikely to be located at high redshift.
We therefore assume that these undetected sources will not  
influence the results and conclusions of this project.

\section{The selection of the candidate sample}

Since only one night of spectroscopy time was available 
for this project, with only a limited part of the sky accessible,
the sample was reduced to objects located at north galactic latitudes,
$b>30^\circ$. In this way the area of sky overlaps with that scanned
by the APM (see below). 
Furthermore, since all the observed radio-loud quasars at $z>4$ have 
radio-spectral
indices clearly flatter than $-0.5$, the spectral index cut-off 
was increased to $\alpha_{1.4-5}>-0.35$ (removing 30\% of the sources). 
In this way, the contamination
of possible `low' redshift galaxies, which are more biased towards steepish 
spectra (Snellen et al. 2001), is reduced.
This resulted in a sample of 3202 candidate objects, selected on their radio 
properties only, over an effective area of $\sim 6400$ deg$^{2}$.

%\subsection*{The APM $-$ POSS-I catalogue}

Optical identification of the resulting sample was carried out using the 
output catalogue of the APM scans of the Palomar Sky Survey (POSS-I) 
photographic plates in the $e$ (red) and $o$ (blue) passbands 
(McMahon \& Irwin 1991). 
The optical radio correlation was done in a similar way as 
for the Jodrell Bank - VLA Astrometric Survey (JVAS), as described in
detail by Snellen et al. (2001). 
 The 3202 sources in the sample were correlated with 
the APM catalogue and were considered to have a candidate optical 
identification if the optical to radio position offset was found to be less 
than $3''$. 
In this way, 2030 sources ($\sim 63 \%$) were found to have an optical
identification. Note that the true percentage of CLASS sources
with optical identifications is slightly higher than this due to 
limitations of the APM catalog, e.g. fragmentation of large galaxies, 
halos of bright stars, and blending of objects such that the centroid
of the merged objects is displaced by a few arcseconds
(see e.g. Snellen et al. 2001). 
Note however, that these problems will mainly
occur for nearby galaxies and bright stars, which will not affect point source
identification statistics of radio sources.

Candidate high redshift quasars were selected using the colour selection
method (Hook et al. 1995), which is based on the assumption that 
objects at high redshift appear redder than those at
low redshift due to intervening Ly$\alpha$ absorption systems shifted into 
the blue passband.
Hook et al. (1996) selected high redshift quasar candidates from the 
JVAS survey with star-like identifications, $e<19.5$, 
and $o-e$ colour $>1.0$.
However,  303 objects in the sample fulfill these selection criteria, 
too many for our limited observing time. Furthermore,
the expected success-rate for finding $z>4$ quasars using these
criteria, is only $0.6\%$ (Hook et al. 1996).
For this project we were only interested in the quasars at 
$z>4$, so we sharpened the selection criteria to 
increase the success-rate and decrease the number
of candidates.

The five previously published radio-selected quasars, have $o-e>2.9$
(see table \ref{litgt4}. We therefore increased 
the colour-cutoff from $o-e>1.0$ to $o-e>2.0$. Although this meant
that our sample is incomplete for $3<z<4$, no effect is expected for
quasars at $z>4$
The main contribution of candidates which turn out not to be
high redshift quasars comes from galaxies at moderate $z\sim0.5$ 
distances, which the APM can not distinguish from 
point sources. The APM morphological classification is particularly crude
towards its magnitude limit. We therefore lowered the magnitude limit
for our candidate sample from $e<19.5$ to $e<19.0$.
The more stringent selection criteria in optical magnitude and 
colour, decreased the number of objects in the target sample from 
303 to 40.

The IPAC/Caltech 
Extragalactic Database (NED) was searched for redshifts of 
the 40 potential target sources.
Seven sources turned out to have measured redshifts, of which three 
galaxies (J1149+2958, z=0.158, Gregg et al. 1996;
J1301+2842, z=0.42, Hook et al. 1998; 
J1436+4129, z=0.17, Hook et al. 1998), and four quasars
(J1258+2909, z=3.51, Hook et al. 1998; J15100+5702, z=4.30,
Hook et al, 1996; GBJ1605+3038, z=2.67;
J1746+6226, z=3.89, Hook et al. 1996).
In addition, the APM, POSS-I, and where available, 
the POSS-II data of the targets, were carefully examined by eye for anomalies 
or possible extendedness. Of the remaining 33 objects, a further 
13 were rejected in this way; 
J1324+3417, J1609+4717, J1625+2017,
J1648+5039, J1710+5028, J0845+4239,
J0900+1357, J0951+3451, J1012+1619,
J1151+6654, J1259+4129, J1605+3029, and J1632+0419.
Hence, 27 objects satisfied our selection criteria, of which 
20 sources were previously unobserved.

\begin{table}
\setlength{\tabcolsep}{1.8mm}
\caption{\label{litgt4} The radio and optical properties of 
 radio-selected quasars at $z>4$, including those discovered by
this work.}
\begin{center}
\begin{tabular}{llllrll}\hline
Name&z& e& o-e&$ S_5$ & $ \alpha_{5-1.4}$& Refs.\\
    & &mag& mag& mJy    &                  &          \\ \hline
J083946+511202  &4.41   &18.5&$>$3.3& 51&+0.11 & 5\\
J091824+063653  &4.19   &18.7&$>$2.7& 36&+0.12 & 5\\
RXJ1028.6$-$0844&4.276  &18.9&\phantom{$>$}2.6 &220&$-$0.16& 1 \\
PKS1251$-$407   &4.460  &20.0&\phantom{$>$}3.7 &240&$-$0.10& 2 \\ 
J132512+112329  &4.40   &18.8&$>$3.2& 62&$-$0.24& 5 \\
GB1428+4217     &4.715  &20.9\dag&$>$3.4\dag&250&$+$0.13& 4 \\ 
GB1508+5714     &4.301  &18.9&$>$3.5&280&$+$0.50& 3,5 \\ 
GB1713+2148     &4.011  &21.0\dag&$>$2.9\dag&330&$-$0.23& 4 \\ \hline
\end{tabular}
\begin{tabular}{l}
\dag R and B-R magnitudes\\
References:\\
1: Zickgraf et al. 1997\\
2: Shaver et al. 1996\\
3: Hook et al. 1995\\
4: Hook et al. 1998\\
5: This paper\\
\end{tabular}
\end{center}
\end{table}

\section{Observations and reduction}

\begin{figure*}
\center{
\psfig{figure=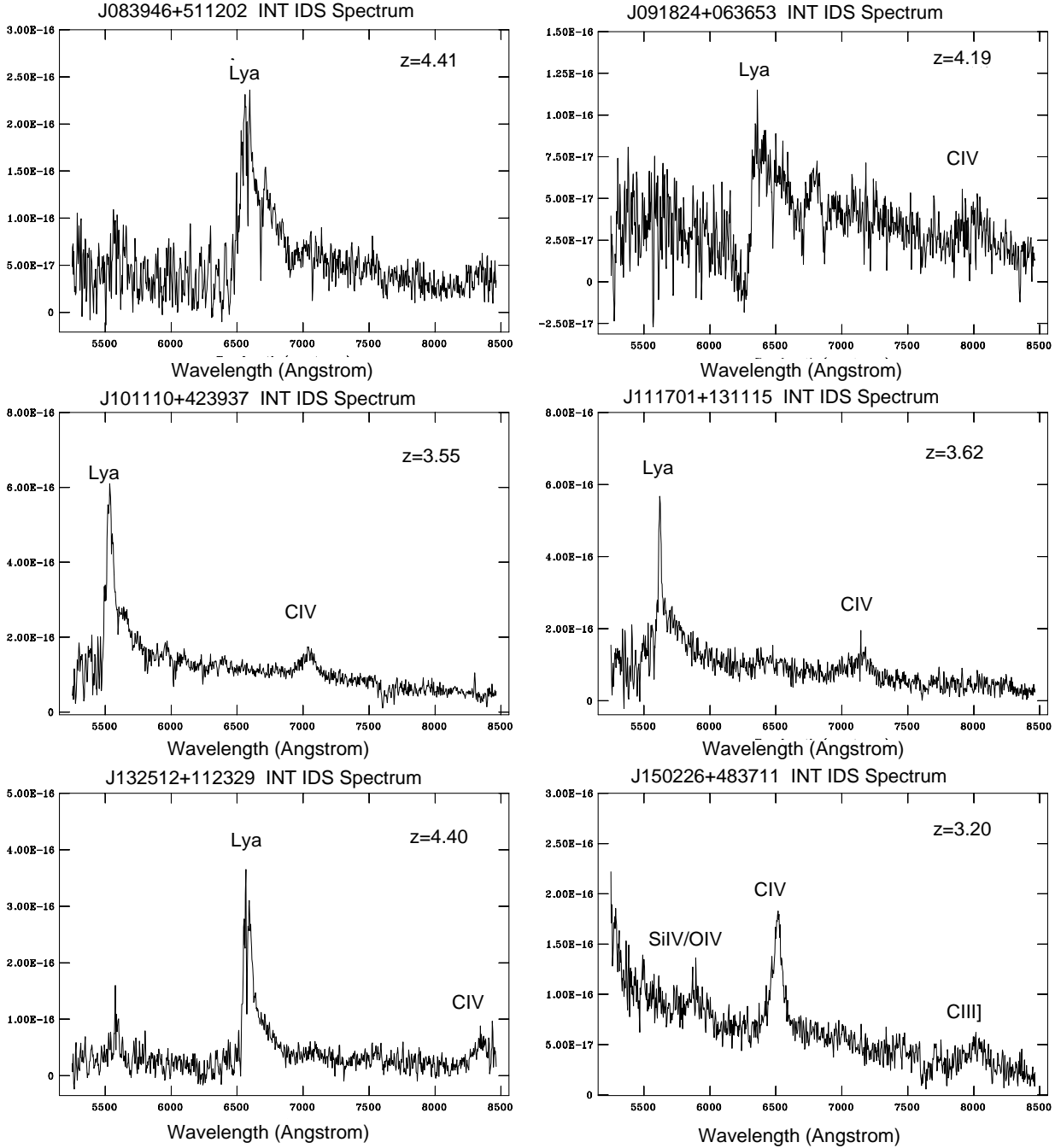,width=17.5cm}}
\caption{\label{spectra} Spectra of the $z>3$ quasars observed by us.}
\end{figure*}

We observed for one night on the 11th of May 1999 in grey time 
with the 2.5m Isaac Newton Telescope (INT), as part of a programme carried 
out in the international CCI 
observing period on the La Palma telescopes in 1999.
The Intermediate Dispersion Spectrograph  (IDS) was used with a Tektronix CCD.
A spectral resolution of $\sim 10 \AA$ was obtained between 5500 and 
8500 \AA, using a slit width of 2 arcseconds and a grating with
300 grooves per mm. Exposure times were 15 minutes for each object.
For efficiency reasons, the orientation of the slit 
was kept constant, which was typically at a parallactic angle 
if the object was observed at zero hour angle.  

The reduction of the spectra was carried out using the `Long Slit' package
of the NOAO's IRAF reduction software. A bias frame was constructed 
by averaging `zero second' exposures taken at the beginning of the night.
This was subtracted from every non-bias frame. The pixel-to-pixel variations
were calibrated using sky flat fields taken at twilight.
Wavelength calibration was carried out by measuring the positions 
on the frames of known emission lines from either a Cu-Ne or Cu-Ar 
calibration lamp. The sky contribution was removed by subtracting a sky
spectrum obtained by fitting a polynomial to the intensities measured along 
the spatial direction outside the vicinity of objects.
One-dimensional spectra were extracted by averaging in the spatial direction 
over an aperture as large as the spatial extent of the brightest emission line.

\begin{table*}
\caption{\label{log} Optical and radio data for objects from the complete 
sample.}
\begin{tabular}{lccccrrcr}\hline
GB6 Name & R.A.        & Decl.&$\Delta r$ & e & o-e & $S_{5}$ & $\alpha_{1.4-5}$ & z \\
         & hh mm ss.ss & dd mm ss.s  &$''$ & &     &  mJy    &                  &   \\ \hline
\multicolumn{2}{l}{Quasars:}\\
\\
J083946+511202 &08 39 46.11 &+51  12  03.0&0.51& 18.45&$>+3.3$&51&+0.11   &4.41\\  
J091824+063653 &09 18 24.38 &+06  36  53.4&0.45& 18.68&$>+2.7$&36&+0.12   &4.19\\     
J101110+423937 &10 11 10.79 &+42  39  37.2&0.62& 18.14&$ +2.5$&47&$-$0.20 &3.55\\       
J111701+131115 &11 17 01.91 &+13  11  15.5&0.88& 18.28&$ +2.3$&45&+0.32   &3.62\\       
J125832+290903$^a$ &12 58 32.21 &+29  09  03.1&0.72& 18.52&$ +2.2$&34&+0.26   &3.47\\
J132512+112329 &13 25 12.52 &+11  23  29.8&0.69& 18.77&$>+3.2$&62&$-$0.24 &4.40\\       
J150226+483711 &15 02 26.89 &+48  37  11.5&0.37& 17.87&$ +2.4$&39&+0.16   &3.20\\       
J151003+570243$^b$ &15 10 02.93 &+57  02  43.4&0.41& 18.88&$>+3.5$&292&+0.50  &4.30\\
J160523+303837$^a$ &16 05 23.70 &+30  38  37.2&2.45& 18.12&  +2.2&56&-0.20   &2.67\\
%J155633+351757 &15 56 33.74 &+35  17  56.9&0.43& 18.68&$ +2.6$&28&$-$0.01 &1.48\\      
%J174442+675047 &17 44 42.18 &+67  50  46.6&0.25& 18.12&$ +2.1$&23&$-$0.29 &3.42\\       
J174614+622654$^b$ &17 46 13.98 &+62  26  54.2&0.14& 18.29&$ +2.6$&589&$-$0.24&3.89\\  \hline
\multicolumn{2}{l}{Galaxies/Featureless:}\\
\\
J103039+580611 &10 30 39.59 &+58  06  10.9&0.34& 18.09&$ +3.1$&80&$-$0.27 & $-$\\       
%J105538+305250 &10 55 38.61 &+30  52  50.7&0.51& 17.83&$ +2.3$&22&+0.36   & $-$\\       
J120542+332147 &12 05 42.85 &+33  21  44.2&2.59& 17.30&$ +2.1$&57&+0.08   & $-$\\       
J121506+534953 &12 15 06.01 &+53  49  53.8&0.09& 18.00&$ +2.3$&54&$-$0.02 & $-$\\       
J125204+134450 &12 52 04.13 &+13  44  49.9&0.98& 18.79&$ +2.1$&52&$-$0.02 & $-$\\
J130125+284254$^a$ &13 01 23.41 &+28  42  53.7&1.03& 18.86&$>+3.3$&35&+0.84   &0.42\\     
J134731+055233 &13 47 31.47 &+05  52  33.4&0.90& 18.54&$ +2.9$&81&$-$0.43 & $-$\\       
J135449+172604 &13 54 49.90 &+17  26  04.0&0.22& 18.66&$>+3.4$&34&+1.59   & $-$\\       
J141842+235816 &14 18 42.80 &+23  58  16.0&0.66& 18.39&$ +3.1$&48&+0.06   & $-$\\       
J145802+465744 &14 58 02.66 &+46  57  45.2&0.65& 18.72&$ +2.6$&31&$-$0.19 & $-$\\       
J153345+060814 &15 33 45.58 &+06  08  14.4&0.63& 18.95&$>+3.2$&49&$-$0.13 & $-$\\       
J154023+614627 &15 40 23.78 &+61  46  26.5&0.63& 18.47&$ +2.7$&49&$-$0.14 & $-$\\       
J164213+244159 &16 42 13.29 &+24  41  58.9&0.38& 18.79&$>+3.0$&30&+0.09   & $-$\\       
J164820+503903 &16 48 20.27 &+50  39  03.6&1.56& 18.80&$>+3.1$&45&$-$0.29 & $-$\\\hline
\multicolumn{2}{l}{Stars:}\\ 
\\
J080131+545419 &08 01 31.54 &+54  54  20.3&1.55& 18.74&$ +2.7$&37&+0.12   &Star\\
J110938+251350 &11 09 38.14 &+25  13  47.9&2.93& 17.07&$ +3.5$&108&$-$0.29&Star\\ \hline      

\multicolumn{2}{l}{$^a$ Observed by Hook et al. 1998}\\
\multicolumn{2}{l}{$^b$ Observed by Hook et al. 1996}\\
\end{tabular}
\end{table*}

\section{Results and Discussion}

The results of the observations are shown in table \ref{log}.
Column 1 gives the J2000 name, columns 2 and 3 give the J2000 right ascension
and declination for the optical APM position. Column 4 gives the optical-radio
position offset in arcseconds. Columns 5 and 6 give the $e$ band magnitude
and $o-e$ optical colour. Column 7 and 8 give the GB6 5GHz flux density
and 1.4$-$5 GHz radio spectral index, and column 9 gives the measured redshift.
For completeness we include the four quasars previously observed
by Hook et al. (1996, 1998).

\begin{figure}
\psfig{figure=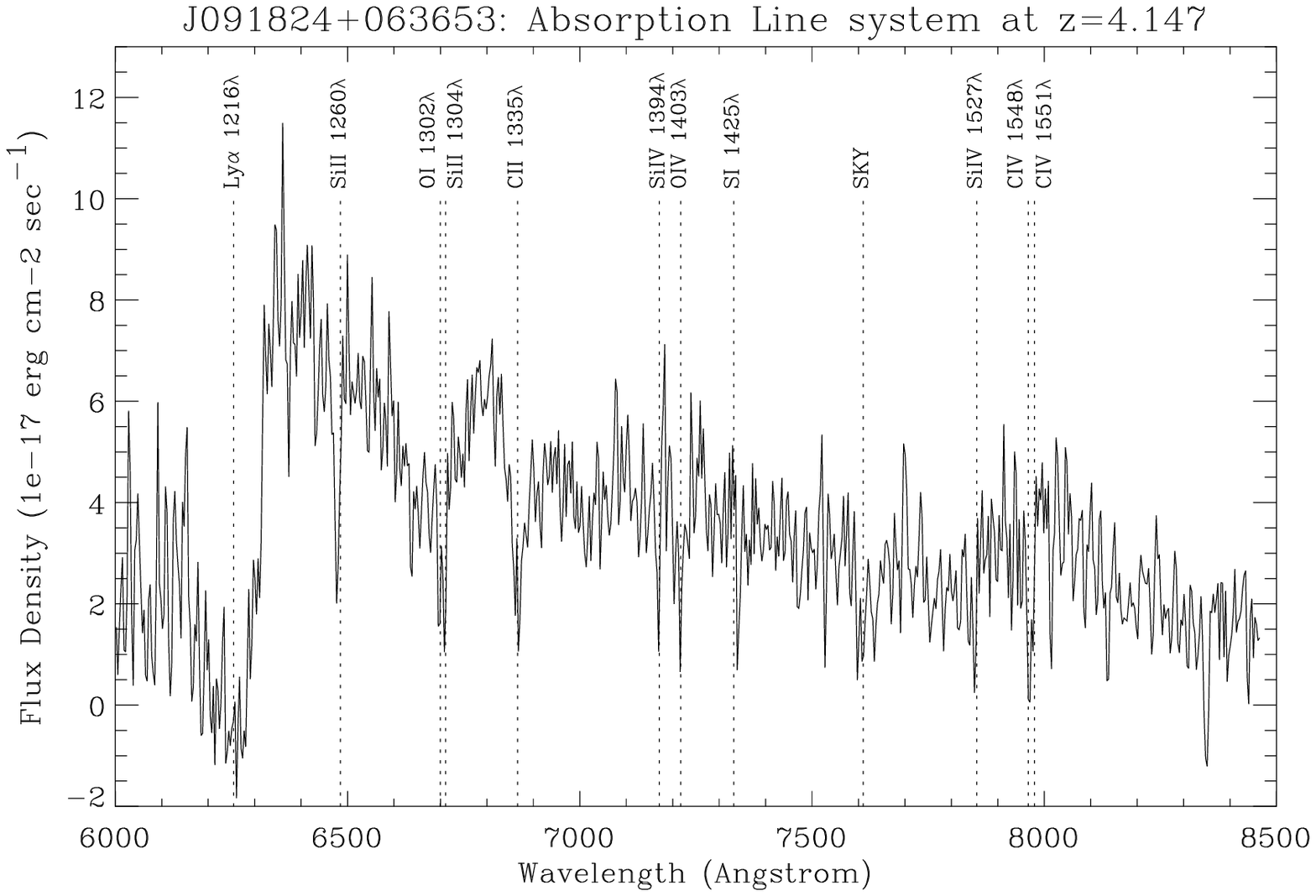,width=8cm}
\caption{\label{abs} Part of the spectrum of J091824+063653 showing the 
absorption line system at z=4.147}
\end{figure}

Observations of the 20 previously unknown sources revealed 6 quasars
(of which 3 are located at $z>4$),
 12 galaxies or featureless objects, and two galactic stars (which are
most likely coincidental foreground objects).
The redshift of all but 2 objects was determined from
Ly$\alpha$1216$\lambda$ and CIV1549$\lambda$ in emission.
The redshifts of J083946+511202 and J132512+112329 were determined from 
the associated Ly$\alpha$1216$\lambda$ absorption in these objects which 
coincide with their broad Ly$\alpha$1216 emission. 
For both objects, the CIV1549$\lambda$
is red-shifted just to the edge of the observed wavelength band, making it 
difficult to determine the redshift for this line. J083946+511202 
shows NV1240$\lambda$ emission at a redshift consistent with the Ly$\alpha$
absorption. The Ly$\alpha$ emission of J091824+063653 is almost 
completely absorbed, and only a very crude redshift can be determined
from the broad CIV1549$\lambda$ of z=4.19.
The strong Ly$\alpha$ absorption, which may
be caused by the host-galaxy,  is accompanied by several absorption
lines at z=4.147 of OI, SI, SiII, OIV, SiIV and CIV (see figure \ref{abs}).

Considering all 27 objects in our final sample 
(including the 7 objects with previously measured redshifts),
the selection criteria used resulted
in a success-rate for finding $z>4$ quasars of 1:7, which is 
a clear improvement over previous studies (1:150, Hook et al. 1996).
Note however, that a negative consequence of this method is a 
lower magnitude limit and incompleteness for quasars located at $3<z<4$.

\begin{figure}
\psfig{figure=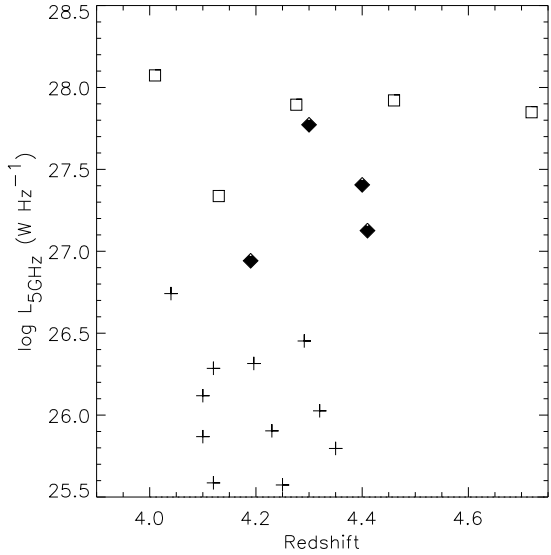,width=8cm}
\caption{\label{lum} The rest-frame 5 GHz radio luminosity of radio-loud 
quasars at $z>4$ from radio-selected samples (squares), optically
selected sampled (crosses; Stern et al. 2000, McMahon et al. priv. comm.), and from the sample 
presented in this paper (diamonds).}
\end{figure}

Figure \ref{lum} shows the rest-frame 5 GHz radio luminosity of radio-loud 
quasars at $z>4$ from radio-selected samples (diamonds), optically
selected sampled (crosses; Stern et al. 2000), and from this sample (boxes).
Where possible, we used the observed 1.4 GHz flux density to compute 
the rest-frame 5 GHz luminosity assuming a flat 
spectral index and H$_0$=50 km sec$^{-1}$ Mpc$^{-1}$ and 
$\Omega_0$=1.
The sample bridges the gap between the strong radio surveys with 
$S_{5GHz}>200$ mJy and the samples of radio-weak quasars that are generated
via radio observations of optically selected quasars.

\begin{figure}
\psfig{figure=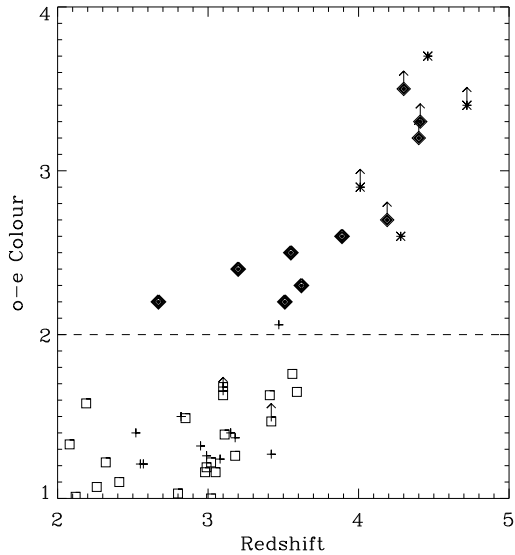,width=8cm}
\caption{\label{colr}
The $o-e$ colours of radio-loud quasars. The diamonds indicate the sources
in our sample. The squares and crosses indicate quasars from Hook et al.
(1996) and (1998) respectively. The stars indicate the remaining previously
known $z>4$ quasars. The dashed line 
indicates the colour-cutoff used for the selection of the sample.} 
\end{figure}

Figure \ref{colr} 
shows the colours of radio-loud quasars as function of 
redshift, for the quasars in our sample (diamonds), of Hook et al. 
(1996-squares; 1998-crosses), and of the remaining $z>4$ quasars 
available in the literature (stars). 
 It shows that the colour-cutoff of $o-e>2.0$, used to select 
the  sample, is unlikely to have caused $z>4$ quasars to be excluded.
We can therefore assume that the 4 $z>4$ quasars are all the radio-loud
quasars with $S_{5GHz}>30$ mJy, and $e<19.0$ in the targeted area of sky. 
This corresponds to a surface density of 1 per $\sim$1600 square
degrees. Note that the use of the POSS-I $e$ plates produces
a strong bias against quasars at $z>4.5$, which is shown in figure 
5, and that we are therefore only targeting objects located at
$4.0<z<4.5$. The surface density we find is about a factor $\sim$ 
15 lower than what is found in a similar 
study targeting radio-quiet quasars at $z>4$ by Storrie-Lombardi 
et al. 
(1998), and implies that the fraction of quasars that are 
very radio-loud  is about 7\%.

\begin{figure}
\psfig{figure=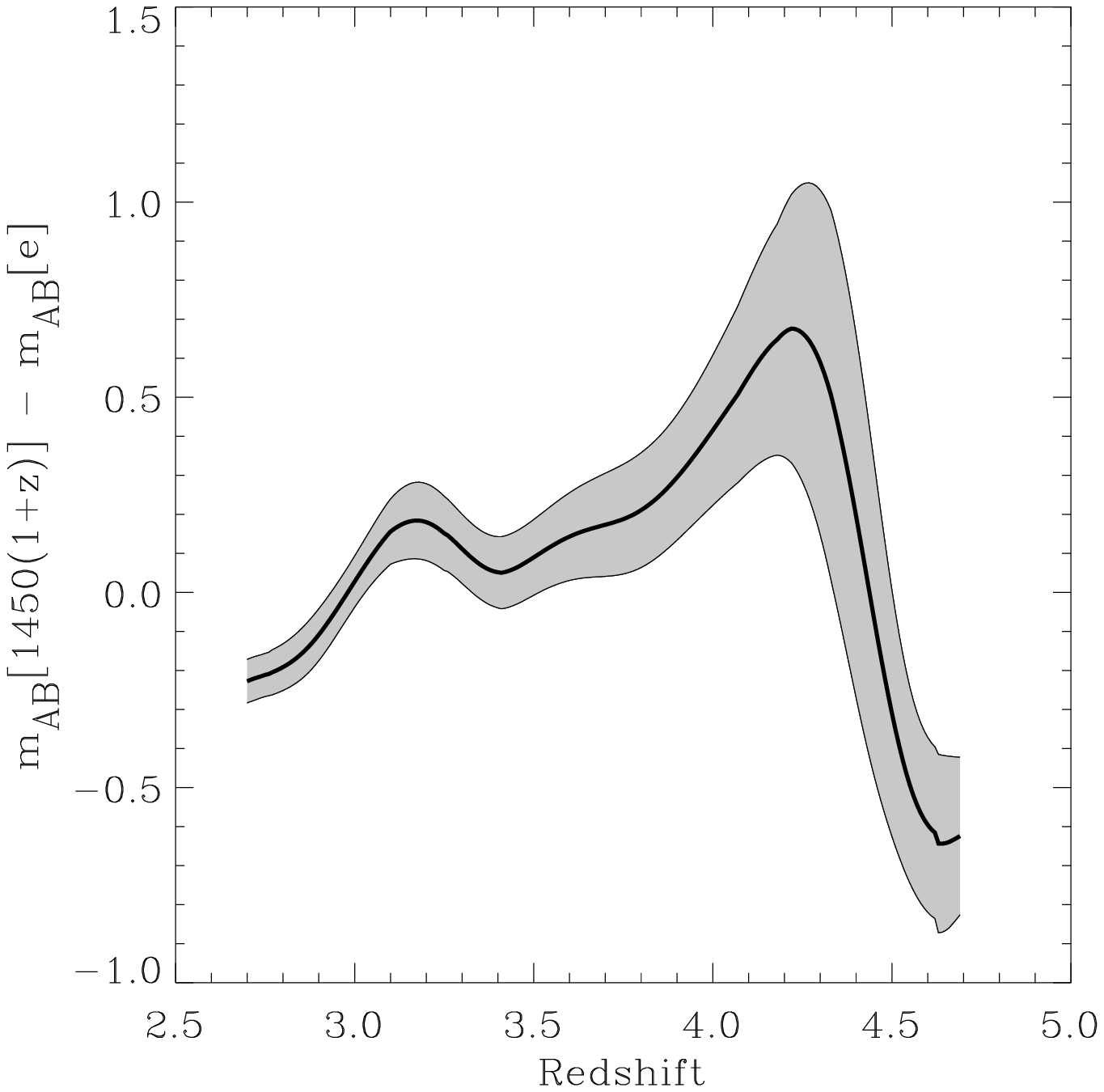,width=8cm}
\caption{\label{qsocol} The K-correction between apparent AB magnitude 
at 1450 (1+z) \AA and apparent AB magnitude through the e filter, plotted
as function as redshift for the quasars in our sample. 
 It is derived from the 
convolution of their optical spectra with the $e$-filter passband. 
The thick line indicates the average, 
and the grey area the scatter.  The two `maxima' are caused the 
Ly$\alpha1216\AA$ and CIV1549$\AA$ emission lines.
The sharp cut-off at $z\sim4.5$ indicates that the use of the POSS-I $e$
plates strongly biases against quasars at $z>4.5$.}
\end{figure}

\section{Summary}

We have reported on the search for distant radio-loud quasars in
the CLASS survey of flat spectrum radio sources. 
Unresolved optical counterparts were selected, as 
measured from APM scans of POSS-I plates, with $e<19.0$ and red
$o-e>2.0$ colours, in an effective area of $\sim6400$ deg$^{2}$. 
Four sources where found to be radio-loud quasars with $z>4$, 
of which one was previously known.
In addition, 4 new quasars at $z> 3$ have been 
found. The selection criteria used result
in a success-rate of $\sim$1:7 for radio-loud quasars at $z>4$, which is 
a significant improvement over previous studies.
It corresponds to a surface density of 1 per 1600 deg$^{2}$, which 
is about a factor of 15 lower than that found in a similar search
for radio-quiet quasars at $z>4$.
The study presented here is 
strongly biased against quasars at $z>4.5$, since the $e$-passband of the 
POSS-I only samples the spectra shortward of 1200$\AA$ at these 
redshifts. 

\section*{Acknowledgements}
We thank the {\it Comit\'e Cient{\'\i}fico International} (CCI) of the IAC
for the allocation of observing time. 
The Isaac Newton Telescope is operated on the island of La Palma by 
the Isaac Newton Group in the Spanish
Observatorio del Roque de los Muchachos of the Instituto de 
Astrof{\'\i}sica de 
Canarias. The NASA/IPAC Extragalactic Database (NED) is 
operated by the Jet Propulsion Laboratory, California Institute of Technology,
 under contract with the National Aeronautics and Space Administration. 
This research was funded by the European Commission under
contract ERBFMRX-CT96-0034 (CERES). RGM thanks the Royal Society for support.
{}
\end{document}